\documentstyle[prb,eqsecnum,aps,epsfig]{revtex}
\begin{document}
\draft

\title{Electronic Structure of Carbon Nanotube Ropes}
\author{A.A. Maarouf, C.L. Kane and E.J. Mele}
\address{Department of Physics \\
Laboratory for Research on the Structure of Matter \\
 University of Pennsylvania\\
Philadelphia, Pennsylvania 19104 }

\date{\today}
\maketitle

\begin{abstract}
We present a tight binding theory to analyze the  motion of
electrons between carbon nanotubes bundled into a carbon nanotube
rope.  The theory is developed starting from a description of the
propagating Bloch waves on ideal tubes, and   the effects of
intertube motion are treated  perturbatively in this basis.
Expressions for the interwall tunneling amplitudes between  states
on neighboring tubes are derived which show the dependence on
chiral angles and intratube crystal momenta. We find that
conservation of crystal momentum along the tube direction suppresses 
interwall coherence in a carbon nanorope containing tubes with
random chiralities. Numerical calculations are presented which
indicate that electronic states in a rope are localized in the
transverse direction with a coherence length corresponding to a
tube diameter.
\end{abstract}

\section{Introduction}

A carbon nanotube is a cylindrical tubule formed by wrapping a graphene
sheet. Single wall carbon nanotubes (SWNTs) can be synthesized in structures $1$ nm in diameter and microns long \cite{Thess}. There has been particular interest in the
 electronic properties of SWNTs which are predicted to exist in
 both conducting and semiconducting forms \cite{Saito}.  Remarkably, it is possible to probe this
 behavior experimentally by contacting individual tubes with lithographically patterned electrodes or by tunneling spectroscopy on single tubes \cite{Kin}.
However,  most methods for synthesizing carbon nanotubes do not produce isolated tubes; instead the tubes self assemble to form a hierarchy of more complex structures.
 At the molecular scale tubes pack together to form bundles or ``ropes" which can
 contain 10-200 tubes. X-ray diffraction reveals that a bundle contains tubes
close packed in a triangular lattice \cite{Thess}, and measurements of the lattice constant and
 tube form factor led initially to the suggestion that these ropes contain primarily
 (10,10) nanotubes, a species predicted to be metallic \cite{Saito}. Subsequent work has demonstrated
 that the ropes likely contain a distribution of tube diameters and chiralities \cite{Rao}. On larger scales
 the ropes bend and entangle, so that the macroscopic morphology of a carbon
 nanotube sample is that of an entangled mat. Carbon nanotubes are also formed in
 various thick multiwalled species which exhibit their own unique electronic behavior \cite{deheer}.

 The electronic properties of isolated SWNTs are controlled by the tube's wrapping vector, curvature and torsion \cite{Kane}. However, in ropes and in multiwalled tubes
 the interactions {\it between} graphene surfaces is expected to play a major role.
 This is the case even  for crystalline graphite in the Bernal structure. Although an  isolated graphene sheet is a zero gap semiconductor, the small residual  interactions between neighboring graphene sheets
 with the ordered $A$-$B$ stacking sequence lead to a small overlap of bands near the Fermi energy and eventually to conducting behavior \cite{graphite}. Graphite is an ordered three dimensional crystal for which  weak intersurface interactions
 are sufficient to establish quantum coherence for electronic states on neighboring sheets.  Thus the electrons can delocalize both parallel to the graphene sheet and perpendicular
 to it.

 Recognizing this, several groups have attempted to estimate the energy scale for
 similar effects in nanotube ropes \cite{Delaney}. Here the situation is much  more delicate, since the
 structure of a (10,10) nanotube does not permit perfect registry between neighboring
 tubes when they are  packed into a triangular lattice. Nevertheless, it is possible to construct
 a nanotube crystal, a hypothesized ordered structure in which each (10,10) tube adopts the
 same orientation, and to study its electronic properties with conventional band theoretic
 methods \cite{Delaney}. Theoretical studies on nanotube ropes show that intertube interactions in a nanotube crystal lead to a mixing of
 forward and backward propagating electronic states near the Fermi energy. The level
 repulsion between these branches leads to suppression or ``pseudogap" in the
 electronic density of states, on an energy scale estimated to be a few tenths of
 an electron volt. It should be noted that these effects are qualitatively different from
 those found in graphite where intersurface interactions lead to band {\it overlap}
 and  thus an enhancement of the Fermi level density of states.

 It has not yet  been possible to extend these ideas to carbon nanotube ropes which
 contain a mixture of tubes with various diameters and chiralities. A direct calculation of
 the electronic structure for such a rope, which we define as ``compositionally disordered," is quite complicated since the system has no translational symmetry either
 along the rope axis or perpendicular to it.

 In this paper we develop a tight binding theory for the coupling between tubes. In the absence of intertube coupling
 the electronic states on an isolated tube are essentially the Bloch waves of the
 graphene sheet wrapped onto the surface of a cylinder and indexed by
 a two dimensional crystal momentum {\bf k}. The essence of our theory
 is to develop the effects of intertube interactions ${\rm t}({\bf k}_1,{\bf k}_2)$ perturbatively. We find that
 the effects of intertube interactions in a disordered rope are quite different from what one obtains for a crystalline rope. In fact,
 we find that compositional disorder introduces an important energy barrier to inter-tube
 hopping within a rope so that intertube coherence is strongly suppressed. We are led
 to conclude that eigenstates in a compositionally disordered rope are strongly localized
 on individual tubes, though they can extend over large distances along the tube direction.
 Numerical results illustrating this effect will be presented in this paper.
 We believe this physics underlies the experimental observation that charge transport at low
 temperature occurs by  hopping conduction in nanotube ropes and mats.

In section II we develop the tunneling model for
 describing the tight binding coupling between neighboring tubes in a rope. In this
 section we derive an analytic expression giving the tunneling amplitude  ${\rm t({\bf k}_1,{\bf k}_2)}$
 between Bloch states on neighboring tubes indexed by momenta ${\rm {\bf k}_1}$ and
 ${\rm {\bf k}_2}$.  In Section III we apply the method to study the electronic structure
 of a rope crystal, and show that the model reproduces well the results of more complete
 band theoretic calculations on this ordered system. In Section IV we then extend the method to
 study the low energy electronic structure in a {\it compositionally disordered} rope and analyze
 the effects of intertube interactions perturbatively. We will also present direct numerical
 calculations on a compositionally disordered rope which probe the transverse localization of
 the electronic states. A brief discussion of the relation of these results to experimental
 data is given in Section V.

\section{Tunneling Model}

In this section we derive an effective tight binding
model which describes the coupling between
the low energy electronic states on neighboring
tubes.  This coupling depends on the chirality
and orientation of the tubes.  Our starting point is
a microscopic tight binding model which describes the
coupling of the carbon $\pi$ orbitals both within a tube
and between tubes,
\begin{equation}
{\cal H} = {\cal H}_0 + {\cal H}_T.
\end{equation}
${\cal H}_0$ is a nearest neighbor tight binding model describing
uncoupled tubes,
\begin{equation}
{\cal H}_0 = - \sum_a \sum_{<ij>} t_\pi c_{ai}^\dagger c_{aj},
\end{equation}
where the index $a$ labels the tubes and $<ij>$ is a sum over
nearest neighbor atoms on each tube.
Tunneling between tubes is also represented by
\begin{equation}
{\cal H}_T = \sum_{<ab>}\sum_{ij} t_{ai,bj} c_{ai}^\dagger c_{bj} + {\rm
H.c.}.
\end{equation}
In the following we will assume that
$t_{ai,bj} = t_{{\bf r}_{ai},{\bf r}_{bj}}$
depends on the positions and relative orientations of the $\pi$
 orbitals on  the $i$ and $j$ atoms.

The eigenstates of ${\cal H}_0$ are plane waves localized in an
individual tube.  Due to the translational symmetry of an
individual tube the eigenstates may be indexed by a tube index
$a$ and a two dimensional momentum ${\bf k}$.  Of course the
periodic boundary conditions imposed by wrapping the graphene
sheet into a cylinder will give a constraint on the possible
values of ${\bf k}$.  In the following, we wish to express the
Hamiltonian in terms of this plane wave basis.  We will focus on
eigenstates with low energy, which have ${\bf k}$ near one of the
corners of the graphite Brillouin zone.

\subsection{Plane Wave Basis}

It is useful to express the eigenstates of the individual
tubes in a basis of plane wave states localized on either
the $A$ or $B$ sublattice.  Let us first focus on a single tube.
The eigenstates may be described by considering a two dimensional
graphene sheet with periodic boundary conditions.
We will find it useful to consider two
coordinate systems for the two dimensional graphene sheet.  As
shown in Fig. 1, the $x$ and $y$ axes are oriented with respect to the
armchair and zig zag axes of the graphene sheet.  The $u$ and $v$
axes, on the other hand are oriented with respect to the tube,
with $u$ pointing down the tube and $v$ pointing around the
circumference.  For armchair tubes these axes coincide, and in
general the angle between the axes is equal to the chiral angle of
the tube.

Suppressing the tube index, for the moment, we let
\begin{equation}
c_i = {1\over \sqrt{N}} \sum_{\bf k}
e^{i {\bf k}\cdot {\bf r}_i} c_{\eta(i){\bf k}},
\end{equation}
where $\eta$ specifies the $A$ or $B$ sublattice
and $N$ is the number of graphite unit cells on the tube.
In this basis, the Hamiltonian for an isolated tube may
be written
\begin{equation}
{\cal H}_0 = - t_\pi \sum_k \gamma_{\bf k} c_{A{\bf k}}^\dagger
c_{B{\bf k}} + {\rm H.c.},
\end{equation}
where
\begin{equation}
\gamma_{\bf k} = \sum_{j=1}^3 e^{i {\bf k}\cdot{\bf d}_j}.
\end{equation}
Here ${\bf d}_j$ are the three nearest neighbor vectors connecting
the $A$ and $B$ sublattice indicated in Fig. 1. At low energy we may focus on the points ${\bf k} = \alpha {\bf K}_\ell + {\bf q}$, where $\alpha = \pm 1$, ${\bf K}_\ell$ are at the corners of the Brillouin zone shown in Fig. 2, and $\ell = -1,0,1$. In the $u-v$ system, the ${\bf K}_\ell$ vectors can be written as
\begin{equation}
\alpha {\bf K}_\ell = \alpha K_0(\cos\omega_\ell,\sin\omega_\ell ),
\end{equation}
where 
\begin{equation}
\omega_\ell = {2\pi \over 3} \ell + \theta
\end{equation} 
is  the angle that the $\ell^{\mbox {th}}$ Fermi vector makes with the $u$ axis.

\begin{figure}
   \epsfxsize=2.5in
   \centerline{\epsffile{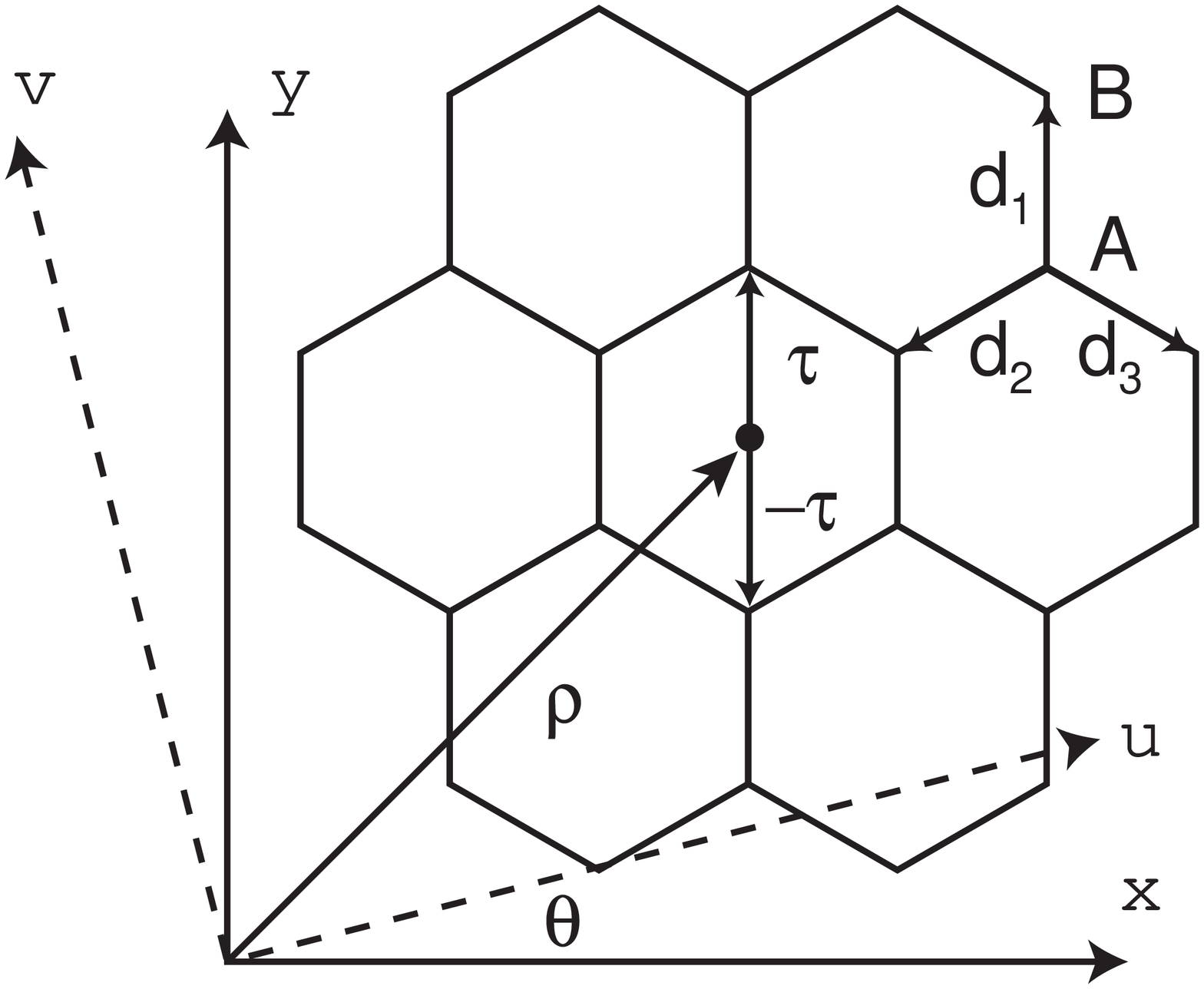}}
   \caption[FIG. 1.]{A graphene sheet. The $x$ and $y$ axes are oriented with respect to the armchair and zigzag axes, while the $u$ and $v$ axes point in the directions of along and around the tube respectively. The vectors ${\bf d}_i$ are the three nearest neighbor vectors connecting the $A$ and $B$ sublattices and ${\bf \rho}$ is the vector point to the center of a hexagon. $\theta$ is the chiral angle.}
\end{figure}

We will now focus on the point ${\bf K}_0$. For small ${\bf q}$, $\gamma_{\alpha{\bf K}_0 + {\bf q}} = -
(\sqrt{3}a/2) (\alpha q_x - i  q_y)$. Introducing a spinor $\psi_{\eta\alpha {\bf q}} = c_{\eta \alpha {\bf K}_0 + {\bf q}}$ the Hamiltonian may then be written,
\begin{equation}
{\cal H}_0 = v \psi_{\alpha {\bf q}}^\dagger ( \alpha q_x \sigma_x
+ q_y \sigma_y ) \psi_{\alpha {\bf q}},
\end{equation}
where $v = \sqrt{3} t_\pi a / 2$ and the $\eta$
indices are suppressed.

The tunneling Hamiltonian may similarly be expressed in this plane
wave basis.
Using (2.4) the term in (2.3) for the bond connecting tubes $a$
and $b$ is
\begin{equation}
{1\over N} \sum_{{\bf r}_a {\bf r}_b} \sum_{{\bf k}_a{\bf k}_b}
t_{{\bf r}_{a}{\bf r}_{b}}
 e^{i( {\bf k}_b\cdot{\bf r}_{b}
 -{\bf k}_a \cdot {\bf r}_{a}) }
 c_{a\eta_a {\bf k}_a}^\dagger  c_{b\eta_b{\bf k}_b},
\end{equation}
where $\eta_{a,b}$ label the sublattice of the lattice site ${\bf
r}_{a,b}$.

\begin{figure}
   \epsfxsize=2.5in
   \centerline{\epsffile{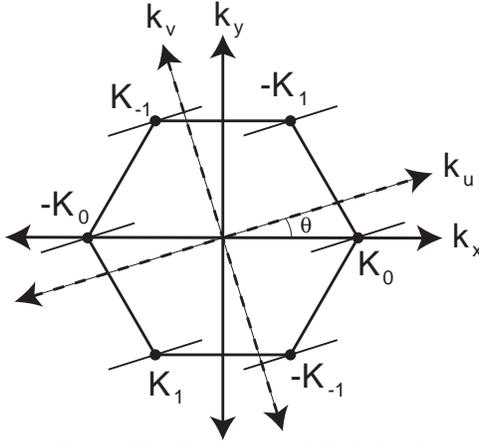}}
   \caption[FIG. 2.]{Brillouin zone of a graphene sheet. ${\bf K}_\ell$, $\ell=-1,0,1$ label the three equivalent Fermi points and $\theta$ is the chiral angle of the tube formed by wrapping the sheet in the $v$ direction.}
\end{figure}

The sums over lattice sites may be evaluated by introducing a
Fourier transform of the tunneling matrix element. As detailed in
Appendix A, we may write
\begin{equation}
{\cal H}_T = \sum_{{\bf G}_a {\bf G}_b} \sum_{ \eta_a \eta_b}
e^{i{\bf G}_a\cdot (\rho_a + \eta_a \tau_a) - {\bf
G}_b\cdot(\rho_b + \eta_b \tau_b)} t_{{\bf k}_a+{\bf G}_a {\bf
k}_b + {\bf G}_b} c_{\eta_a{\bf k}_a}^\dagger  c_{\eta_b{\bf
k}_b},
\end{equation}
where
\begin{equation}
t_{{\bf k}_a,{\bf k}_b} = {1\over{N A_{\rm cell}^2}}
\int d^2 r_a d^2 r_b t({\bf r}_a,{\bf r}_b)
e^{-i{\bf k}_a\cdot{\bf r}_a + i {\bf k}_b\cdot{\bf r}_b },
\end{equation}
and $G$ is a reciprocal lattice vector.

We now specialize to eigenstates in the vicinity of the Fermi
points, ${\bf k} = \alpha{\bf K}_0 + {\bf q}$. We may express the
sum over the $G$'s as a sum over equivalent ${\bf
K}$ points which are related to ${\bf K}_0$ by a reciprocal lattice vector.  In the following we will see that this sum is
dominated by the ${\bf K}_0$, ${\bf K}_1$ and ${\bf K}_{-1}$, which
lie in the ``first star" in reciprocal space.  Since ${\bf
K}_0\cdot{\bf \tau} = 0$, the sum becomes
\begin{equation}
{\cal H}_T = \sum_{\alpha_a\eta_a\alpha_b\eta_b{\bf q}_a {\bf
q}_b}
T(\alpha_a\eta_a{\bf q}_a | \alpha_b\eta_b {\bf q}_b)
\psi_{a\alpha_a\eta_a{\bf q}_a}^\dagger
\psi_{b\alpha_b\eta_b{\bf q}_b},
\end{equation}
with
\begin{equation}
T(\alpha_a\eta_a{\bf q}_a | \alpha_b\eta_b {\bf q}_b)
=
\sum_{\ell_a \ell_b=-1}^1 e^{ i \alpha_a {\bf K}_{a \ell_a} \cdot (\rho_a + \eta_a
\tau_a) - i \alpha_b {\bf K}_{b \ell_b} \cdot (\rho_b + \eta_b \tau_b)}
t_{\alpha_a {\bf K}_{a \ell_a}+ {\bf q}_a, \alpha_b{\bf K}_{b \ell_b} + {\bf
q}_b}.
\end{equation}

We shall also find it useful to express the tunneling Hamiltonian
in a basis in which the bare Hamiltonian describing the tubes is
diagonal.  This is accomplished by performing a rotation in the
sublattice index space to make (2.9) diagonal. Specifically, for a
tube with chiral angle $\theta$, the eigenstates will have
momentum ${\bf k} = \alpha {\bf K}_0 + {\bf q}$, with $(q_x,q_y) =
q (\cos\theta,\sin\theta)$.  Equation (2.9) is then
\begin{equation}
{\cal H}_0 = v \psi^\dagger_{\alpha q} \alpha q (
e^{-i\alpha\theta} \sigma^+ + e^{i\alpha\theta} \sigma^-)
\psi_{\alpha q}.
\end{equation}
Using the transformation $\psi_{\alpha q} = U(\alpha,\theta)
\psi'_{\alpha q}$, with
\begin{equation}
U(\alpha,\theta) =   e^{-i  {1\over 2}\alpha \theta \sigma^z}
e^{-i{\pi\over 4}\alpha \sigma^y},
\end{equation}
the Hamiltonian becomes
\begin{equation}
{\cal H}_0 = v q ( \psi'^\dagger_{\alpha q R} \psi'_{\alpha q R} -
\psi'^\dagger_{\alpha q L} \psi'_{\alpha q L}).
\end{equation}
In the $(R,L)$ basis, the tunneling matrix has the form
\begin{equation}
T'(\alpha_a\eta_a'{\bf q}_a | \alpha_b\eta_b' {\bf q}_b) =
U^\dagger(\alpha_a,\theta_a)_{\eta_a'\eta_a}
 T(\alpha_a\eta_a{\bf q}_a | \alpha_b\eta_b {\bf q}_b)
U(\alpha_b,\theta_b)_{\eta_b\eta_b'},
\end{equation}
which may be written as
\begin{equation}
T'(\alpha_a\eta_a'{\bf q}_a | \alpha_b\eta_b' {\bf q}_b)
 =\sum_{\ell_a \ell_b=-1}^1 e^{ i \alpha_a {\bf K}_{a \ell_a} \cdot
\rho_a - i \alpha_b {\bf K}_{b \ell_b} \cdot \rho_b }t_{\alpha_a {\bf K}_{a \ell_a}+ {\bf q}_a, \alpha_b{\bf
K}_{b \ell_b} + {\bf q}_b} M_{\eta_a'\eta_b'},
\end{equation}
where
\begin{equation}
M =
 {1\over 2} \left[
        \begin{array}{*2c}
          f_{\alpha_{a}}^{\ell_a}   f_{\alpha_{b}}^{\ell_b \ast}
      &  f_{\alpha_{a}}^{\ell_a}   f_{-\alpha_{b}}^{\ell_b \ast} \\
 f_{-\alpha_{a}}^{\ell_a}   f_{\alpha_{b}}^{\ell_b \ast}
      &  f_{-\alpha_{a}}^{\ell_a}   f_{-\alpha_{b}}^{\ell_b \ast}
        \end{array} \right],
\end{equation}
\begin{equation}
f_{\alpha}^{\ell} = e^{i \phi_\ell} + \alpha e^{- i \phi_\ell},
\end{equation}
and
\begin{equation}
\phi_\ell = {1 \over 2}(2\pi - \omega_\ell).
\end{equation}

\subsection{Tunneling Matrix Elements}

For simplicity, we suppose that the matrix elements $t_{ij}$ for
tunneling
between atoms on different tubes depend
only on the distance between the atoms and are of the form,
\begin{equation}
t_{ij} = t_0 e^{-d_{ij}/a_0},
\end{equation}
where $d_{ij}$ is the distance between atoms $i$ and $j$.

It is useful to introduce two dimensional coordinates which are
oriented relative to the tube's axis.  Let us define a two
dimensional vector ${\bf r} = (u,v)$, where $u$ is the distance
down the tube axis, and $v$ is the distance around the tube
measured from the ``contact line" as shown in Fig. 3.

\begin{figure}
   \epsfxsize=2.5in
   \centerline{\epsffile{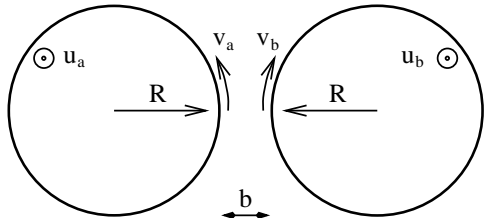}}
    \caption[FIG. 3.]{cross section of two parallel tubes of radius $R$, and with a separation $b$.}
\end{figure}

Suppose the two tubes have a separation $b$ as shown in Fig. 3.
Then, the distance  is given by
\begin{eqnarray}
d({\bf r}_a,{\bf r}_b)^2  = & (u_a-u_b)^2 +
(R\sin {v_a\over R} - R\sin {v_b\over R})^2\\
&\nonumber + (b + 2R - R\cos{v_a\over R} - R\cos{v_b\over R})^2.
\end{eqnarray}
Since the range of the tunneling interaction $a_0$ is of order
$.5$\AA, while $R \approx 7$\AA
and $b\approx$ 3.4\AA,
it is useful to expand (2.24) for $u$,$v \ll b$,$R$,
\begin{equation}
d({\bf r}_a,{\bf r}_b) = b + {|{\bf r}_a - {\bf r}_b|^2\over {2b}} +
{v_a^2 + v_b^2\over {2R}}.
\end{equation}
It follows
that the tunneling matrix element has a Gaussian dependence
on ${\bf r}_a$ and ${\bf r}_b$,
\begin{equation}
t({\bf r}_a,{\bf r}_b) = t_0 e^{-b/a_0}
 e^{-{1\over 2}[{|{\bf r}_a-{\bf r}_b|^2\over {b a_0}}
              + {v_a^2 + v_b^2\over {R a_0}}] }.
\end{equation}
We may now use (2.26) to evaluate the Fourier transform of the
matrix elements.
The Gaussian form allows this to be done simply.  Using the
fact that the total area of the graphene sheet is given by
$N A_{\rm cell} = 2\pi R L$, we find
\begin{equation}
t_{{\bf k}_a{\bf k}_b} =
{2\pi b a_0\over{A_{\rm cell}}}
t_0
e^{- b/a_0}
e^{-{b a_0\over 4} (|{\bf k}_a|^2 + |{\bf k}_b|^2)}
\sqrt{a_0\over{4 \pi R}}
e^{-{R a_0\over 4} (k_{a v} - k_{b v})^2}
 \delta_{k_{a u}k_{b u}}.
\end{equation}
We now use (2.27) to evaluate the low energy tunneling matrix
elements.  Due to the exponential dependence on $|{\bf k}_a|^2$
the sum on ${\bf K}_a$ in (2.11) will be
dominated by three terms ${\bf K}_{ai}$ in the ``first star"
in which $|{\bf K}| = K_0 = 4\pi/(3a)$.
Specifically, estimating the parameters $a = 2.5$ \AA,
$b = 3.4$ {\AA} and $a_0 =0.5$ \AA, the exponent of the
last term is approximately $8\pi^2 b a_0/9a^2 \approx 2.4$.
Thus the next star at $\sqrt{3} K_0$ will be suppressed by a
factor of $\exp(- 2(2.4)) = 0.01$, justifying the first star
approximation.
For ${\bf k}_{a(b)} = \alpha_{a(b)} {\bf K}_{ai(bj)} + {\bf
q}_{a(b)}$ we then have
\begin{equation}
t_{{\bf k}_a{\bf k}_b} = t_T \delta_{k_{a u} k_{b u}} e^{-{1\over
4} R a_0 (k_{a v} - k_{b v})^2},
\end{equation}
with
\begin{equation}
t_T = {2\pi b a_0\over{A_{\rm cell}}}
e^{- b/a_0}
e^{- {1\over 2} b a_0 K_0^2}
 t_0.
\end{equation}
Using (2.27) we then arrive at a final expression for the tunneling
matrix element relating eigenstates on two tubes.
\begin{eqnarray}
T(\alpha_a\eta_a{\bf q}_a | \alpha_b\eta_a {\bf q}_b) 
=  t_T \sum_{\ell_a \ell_b = -1}^1
e^{ i \alpha_a {\bf K}_{a \ell_a} \cdot
(\rho_a + \eta_a \tau_a) - i \alpha_b {\bf K}_{b \ell_b} \cdot
(\rho_b + \eta_b \tau_b)}
\delta_{k_{a u},k_{b u}}
 \nonumber \\
  \times e^{-{1\over 4} R a_0 (k_{a v} - k_{b v})^2} |_{{\bf k}_{a(b)} =
\alpha_{a(b)} {\bf K}_{a \ell_a (b \ell_b)} + {\bf q}_{a(b)}}. 
\end{eqnarray}

\subsection{Estimate of $t_T$ from the band structure of graphite}

The tunneling model described above may be used to describe the
coupling between flat graphene sheets.  Since the transverse
bandwidth of graphite is well known, this allows us to estimate
the prefactor $t_T$ in the tunneling matrix element.  The coupling
between two flat graphene sheets is described by the $R \rightarrow \infty$
limit of the above theory.  In this case, the Gaussian dependence
on $k_{a v} - k_{b v}$ can be written as a (kronecker) delta
function:
\begin{equation}
\sqrt{a_0\over{4\pi R}} e^{-{1\over 4} R a_0 (k_{a v} - k_{b
v})^2} \rightarrow \delta_{k_{av}k_{bv}}.
\end{equation}
We thus obtain
\begin{equation}
t_{{\bf k}_a{\bf k}_b} = t_G \delta_{{\bf k}_a{\bf k}_b},
\end{equation}
with
\begin{equation}
t_G = {2\pi b a_0 \over A_{\rm cell}} t_0 e^{- b/a_0}
e^{- {1\over 2}b a_0 K_0^2}.
\end{equation}

For this calculation, we find it most convenient to use the
sublattice basis for the electronic eigenstates.  Using (2.27) the
tunneling Hamiltonian for two graphene sheets is then,
\begin{equation}
{\cal H}_T = \sum_{{\bf q}\alpha\eta_a\eta_b}
T(\alpha \eta_a {\bf q}| \alpha \eta_b {\bf q})
\psi^\dagger_{a\alpha\eta_a{\bf q}}
\psi_{b\alpha\eta_b{\bf q}},
\end{equation}
where
\begin{equation}
T(\alpha \eta_a {\bf q}| \alpha \eta_b {\bf q}) = t_G
\sum_{\ell=-1}^1 e^{i\alpha{\bf K}_\ell \cdot(\Delta{\bf \rho} +
(\eta_a-\eta_b){\bf \tau} )},
\end{equation}
and $\Delta{\bf \rho} = {\bf \rho}_a - {\bf \rho}_b$.
For $AB$ stacking of graphite, $\Delta{\bf\rho} = {\bf \tau}$,
so, using $\sum_{\bf K} \exp i\alpha {\bf K}\cdot{\bf \tau} = 0$ the
only nonzero term is
\begin{equation}
T(\alpha 1 {\bf q}| \alpha -1 {\bf q}) = 3 t_G.
\end{equation}
Thus tunneling only connects the $A$ sublattice on the "A" sheet to
the $B$ sublattice on the "B" sheet.  This is to be expected, since
an atom on the $B$ sublattice of the "A" sheet sits above a hexagon
on the "B" sheet.  The Hamiltonian for an $AB$ stacked crystal of
graphene planes then has the form,
\begin{equation}
{\cal H} = \sum_s v \psi_{s\alpha {\bf q}}^\dagger ( \alpha q_x
\sigma_x + q_y \sigma_y) \psi_{s\alpha {\bf q}} +  3 t_G \sum_{s}
\psi_{2s\alpha A{\bf q}}^\dagger \psi_{2s+1 \alpha B {\bf q}} +
\psi_{2s\alpha A{\bf q}}^\dagger \psi_{2s-1 \alpha B {\bf q}},
\end{equation}
where $s$ indexes the graphene sheets.  This can be simplified by
introducing a transformation which interchanges the $A$ and $B$
sublattices of the graphene lattice when $\alpha = -1$, followed
by a transformation which interchanges the $A$ and $B$ sublattice on
the odd $(2s\pm 1)$ graphene layers.  The Hamiltonian then has the
simpler form,
\begin{equation}
{\cal H} = \sum_s v \psi^\dagger_s {\bf q}\cdot{\bf \sigma} \psi_s
+ \sum_{<ss'>} {3\over 2} t_G
 \psi^\dagger_s (1+\alpha\sigma_z)
\psi_{s'}.
\end{equation}
This leads to an energy dispersion
\begin{equation}
E({\bf q},q_z) = 3 t_G \cos b q_z
\pm \sqrt{ v^2 |{\bf q}|^2 + (3 t_G \cos b q_z)^2}.
\end{equation}
The bandwidth for transverse motion is then
$W = 12 t_G$.  Experimentally, the bandwidth of graphite
is  in the range $W = 1.2-1.6$ eV \cite{graphite}.  This leads to an estimate $t_G
= 0.1$ eV.

Comparing
(2.29) and (2.33) we may relate the tube tunneling matrix
element to that of graphite,
\begin{equation}
t_T = \sqrt{a_0\over{4\pi R}} t_G.
\end{equation}
For a tube with radius $R = 7${\AA} we find
\begin{equation}
t_T = 7.5 {\mbox{meV}}.
\end{equation}

\section{Electronic Structure of a Rope Crystal}

We now apply the tunneling model described above to the problem of
the electronic structure of nanotube ropes.  We begin by
considering the simpler problem of an orientationally ordered
crystal of (10,10) tubes.  We then consider a compositionally
disordered rope.

(10,10) tubes can be arranged in a triangular lattice in which
each tube has the same orientation, and the tubes face each other
via $A$-$A$ coupling, $B$-$B$ coupling and hexagon-hexagon coupling. For
tunneling in the same subband, $k_{av} = k_{bv}$. Again, we find
it useful to use the sublattice basis for the tube eigenstates.
For each bond, the tunneling between the pair of tubes is
described by equation (2.27) with
\begin{equation}
T(\alpha \eta_a {\bf q}| \alpha \eta_b {\bf q}) =
t_T
\sum_{\ell=-1}^1 e^{i\alpha{\bf K}_\ell \cdot(\Delta{\bf \rho}
+ (\eta_a+\eta_b){\bf \tau} )}.
\end{equation}
For an $A$-$A$ bond it is only nonzero for $\eta_a = \eta_b = 1$.
Using a matrix notation for the $\eta$ indices:
\begin{equation}
T^{AA}(\alpha \eta_a {\bf q}| \alpha \eta_b {\bf q}) = {3\over 2} t_T (1
+ \sigma_z).
\end{equation}
Similarly, for a $B$-$B$ bond,
\begin{equation}
T^{BB}(\alpha \eta_a {\bf q}| \alpha \eta_b {\bf q}) = {3\over 2} t_T
(1-\sigma_z).
\end{equation}
For a hexagon-hexagon bond, we have $\Delta\rho = 0$,
so
\begin{equation}
T^{AB}(\alpha \eta_a {\bf q}| \alpha \eta_b {\bf q})= 3 t_T \sigma_x.
\end{equation}
The Hamiltonian describing the transverse motion will then have the
form,
\begin{equation}
{\cal H}_T = \sum_{\bf q} (\gamma_1 + \gamma_2)
+ \sigma_z (\gamma_1 - \gamma_2) + 2\sigma_x \gamma_3,
\end{equation}
where
\begin{equation}
\gamma_i = 6 t_T \cos {\bf q}\cdot{\bf a}_i,
\end{equation}
where ${\bf a}_i$ are the three nearest neighbor
vectors in the triangular tube lattice.
Then,
\begin{equation}
E(q_x,{\bf q}) = \gamma_1 + \gamma_2
\pm \sqrt{(v q_x - 2 \gamma_3)^2 + (\gamma_1 - \gamma_2)^2}.
\end{equation}
From the above estimate of $t_T = 7.5$ meV, the density of states is plotted in Fig. 4, showing
a pseudogap feature, associated with an energy of
order $12 t_T \approx .09$ eV.  The energy scale of this pseudogap
agrees well with the results of more sophisticated electronic
structure calculations \cite{Delaney}.

\begin{figure}
   \epsfxsize=2.5in
   \centerline{\epsffile{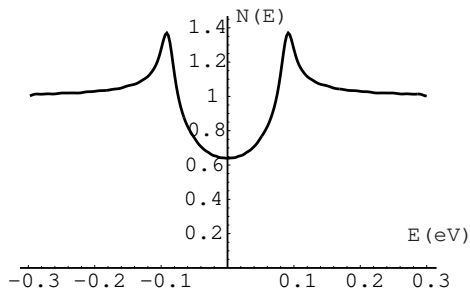}}
   \caption[FIG. 4.]{Density of states of a $(10,10)$ crystal. A pseudogap of about $0.09$ eV develops at the Fermi level.}	
\end{figure}

\section{Compositional Disorder}

A compositionally disordered rope contains a random distribution
of chiral tubes.  Since tubes with different chiralities have
different periodicities, Bloch's theorem is of little use for
describing the eigenstates of the entire rope.  Nonetheless, in
the absence of coupling between the tubes, we know that the
eigenstates on each tube are plane waves.  Our approach is to
describe the coupling between these plane waves perturbatively. We
begin by considering the simpler problem of the electronic
structure of two coupled nanotubes of different chirality.

\subsection{Coupling between two tubes of different chirality}

The electronic coupling between two tubes of different chirality
conserves the momentum along the tube up to reciprocal lattice
vectors in either tube.  As we have argued in section II, the sum
over reciprocal lattice vectors is dominated by the terms in which
${\bf k} + {\bf G}$ are near the first star of ${\bf K}$ points.
When the coupling between the tubes is weak, it is useful to view
this as {\it momentum conserving} coupling between states located
near the three equivalent ${\bf K}$ points.  It must be kept in
mind that two states ${\bf K}_0 + {\bf q}$ and ${\bf K}_1 + {\bf
q}$ in the vicinity of different ${\bf K}$'s are actually the same
state.

Fig. 5 shows the Brillouin zones of two nanotubes with chiral
angles $\theta_a$ and $\theta_b$ oriented so that the $u_a$ and $u_b$ axes
coincide.  Since the tunneling Hamiltonian conserves $k_u$, it
is convenient to view the band structure of the pair of tubes as a
function of $k_u$.  Consider first the band structure in the
absence of coupling.  The solid bands describe the
low energy states on one tube, while the dotted ones show the
states on the other tube. Each set of bands is replicated three
times, reflecting the three equivalent ${\bf K}$ points.  In Fig. 6 we show the band structure in the vicinity of the Fermi points with the minimum momentum mismatch, for the uncoupled system(Fig. 6a), and for the couple one(Fig. 6b). To
lowest order, the effect of the coupling is only important near
points of degeneracy, i.e. where we have band crossing. This
occurs in two cases. The first, which we refer to as a
``backscattering gap" occurs when the right and left moving bands
on each tube cross. The second, which we refer to as a ``tunneling
gap"  occurs when the left moving band on one tube crosses the
right moving band on the other tube.

\begin{figure}
   \epsfxsize=2.3in
   \centerline{\epsffile{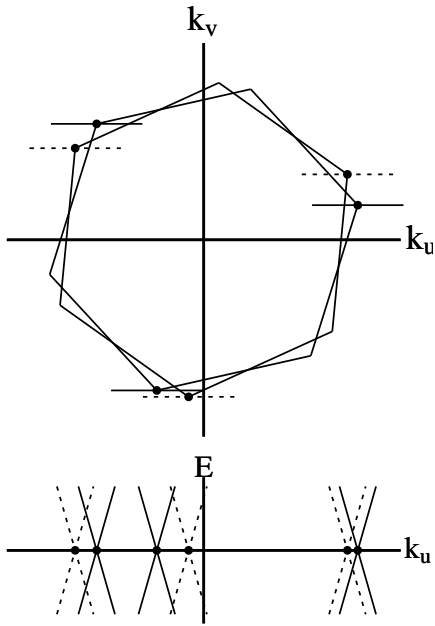}} 
   \caption[FIG. 5.]{Brillouin zones of two tubes of different chiralities. The zones are rotated  such that the $u_a$ and $u_b$ axes coincide. In the lower part of the figure we show the low energy band structure of the two metallic tubes. Solid bands belong to one tube while dotted ones belong to the other one. The bands are replicated three times, reflecting the three equivalent $K$ points. }
\end{figure}

The effect of the coupling depends on two crucial energy scales:
(1) the tunneling matrix element, $t$, which we estimated in
(2.27) to be less than 7.5 meV, and (2) the energy mismatch
\begin{equation}
\Delta E = v [\alpha_a ({\bf K}_{a \ell_a})_u - \alpha_b ({\bf K}_{b \ell_b})_u ],
\end{equation}
which determines the energy at which the right and left moving
bands on tubes $a$ and $b$ cross.  In general, this energy mismatch
depends on the chiral angles $\theta_a$ and $\theta_b$ of the two
tubes as well as the Fermi point indices $i$ and $j$.  In Fig. 7 we
show the variation of the energy mismatch $\Delta E$ with the tube
chirality for all the metallic tubes with diameters between $1.2-1.5$ nm. Tubes which are mirror images to one another have the same diameter and energy difference. The offset of the $u$ momentum is taken at a zigzag
tube; in that case an (18,0) tube. As we see the typical energy mismatch is a few hundred meV. The
fact that $t \ll \Delta E$ simplifies the problem considerably and
justifies our perturbative approach.  Of course it breaks down in
special cases when $\Delta E$ is zero or very small, which occurs,
for instance when the two tubes are mirror images of each other.

\begin{figure}
   \epsfxsize=4.0in
   \centerline{\epsffile{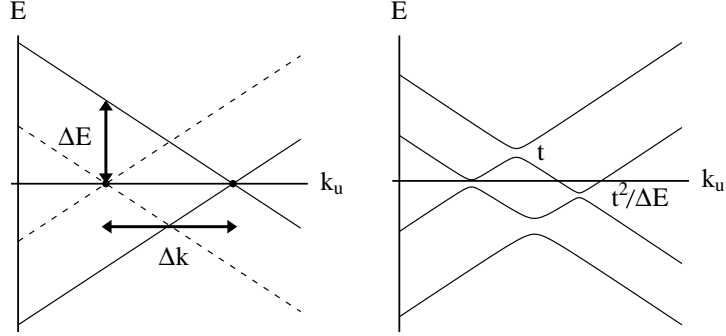}}
   \caption[FIG. 6.]{(a) Band structure of the two(uncoupled) tubes in the vicinity of the Fermi point with the minimum momentum mismatch $\Delta k$, which defines the important energy scale  $\Delta E = v \Delta k$. (b) Band structure of the coupled tubes near the Fermi energy. Backscattering gaps open quadratically with the tunneling strength $t$, whereas tunneling gaps are linear in $t$. In general, backscattering gaps open at different energies.}
\end{figure}

\begin{figure}
\epsfxsize=3.0in
\centerline{\epsffile{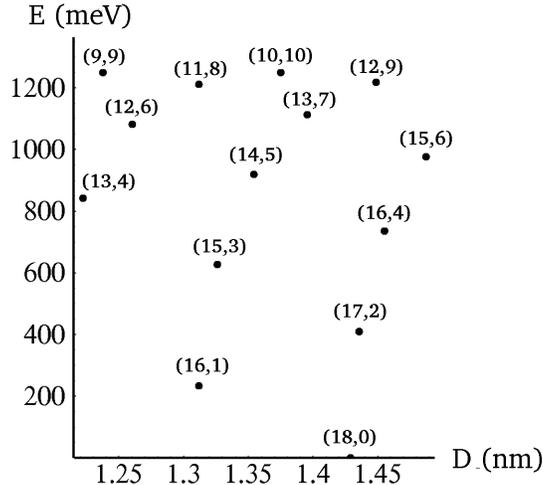}}
\caption[FIG. 7.]{A plot showing the minimum energy mismatch $\Delta E$ for different metallic tubes with radii lying between $1.2$ and $1.5$ nm. The energy offset is taken at a zigzag tube$(18,0)$}
\end{figure}

The tunneling Hamiltonian couples left and right movers in first
order, and therefore the tunneling gap is linear in the tunneling
strength $t$. The gap opens at $\Delta E/2$ and since $t$ is much
smaller than $\Delta E$, one concludes that tube-tube coupling has
a small effect near the Fermi energy.

Backscattering gaps form through second order coupling between
left and right movers on the same tube, and hence the gap is
second order in the tunneling strength; $E_g \sim t^2 / \Delta E$,
which is of order $1$ meV. Furthermore, because the tunneling
matrix elements are not invariant under the interchange of the two
tubes, the backscattering gap of each tube opens at a different
energy. This means that the effect on the density of states near
the Fermi energy is weakened by the wiggling of the gaps. In what
follows, we present quantitative arguments justifying our
expectations.

We focus first on the tunneling gaps. The movers on each tube
couple in first order. In general, crossing occurs at three
different energies, as dictated by the momentum mismatch between
the Fermi points of the two tubes(see Fig. 5). Since we are ultimately interested in the effect of tube interactions on the states nearest to the Fermi level, we only consider points with the lowest lying crossing. We denote these by $\alpha_a {\bf K}_{ai}$ and $\alpha_b {\bf K}_{bj}$.
To find the magnitude of the gaps and their offsets we need to
diagonalize the first order perturbation matrix.  We thus consider
the Hamiltonian which couples the right moving states on tube $a$
with the left moving states on tube $b$, and we diagonalize the
matrix
\begin{equation}
 {\cal H}^{T}_{a} = \left[ \begin{array}{*2c} v({\bf q}-\alpha_a {\bf K}_{ai})_u  & T'(\alpha_a R {\bf q}_a | \alpha_b L {\bf q}_b) \\ T'^*(\alpha_a R {\bf q}_a | \alpha_b L {\bf q}_b) & -v({\bf q}-\alpha_b {\bf K}_{bj})_u \\ \end{array} \right],
\end{equation}
where ${\bf q}$ is measured from the Fermi point of tube $a$. For $\alpha_a, \alpha_b =+1$ the magnitude of the tunneling gap is
\begin{equation}
E^{T}_{g} = 2|T'_{RL}|= 4 t_T e^{-{1\over 4} R a_0 K_0^2 (\sin\omega_{a i} +  \sin\omega_{b j} )^2} \Big | \cos{\omega_{a i} \over 2}  \sin{\omega_{b j} \over 2} \Big |.
\end{equation}
It is centered about an energy $ E^{T}_{o}= \Delta E/2 $ above
the Fermi energy.  Notice that the $+$ sign in the argument of the exponential is due to the fact that the tubes face each other from the outside. As we see from Fig. 7, the average energy separation $\Delta E \sim 300$ meV, whereas the tunneling gap $E^{T}_{g} < 30$ meV. This means that in a rope with
a random distribution of chiralities, the opening of such gaps
will have a negligible effect near the Fermi level.

Now we focus on the backscattering gaps which form near the Fermi
level. Since the left and right moving states on the same tube do
not couple in first order, we
 use second order degenerate perturbation theory. We are interested
in calculating the size of the resulting gap as well as the offset
of the gap. In order to calculate these, we need to diagonalize
the matrix which arises from the second order coupling between the
states.  In general, we have tunneling between all Fermi points on
each tube. Since the magnitude of the backscattering gaps varies quadratically with the tunneling strength and inversely with the
energy difference $\Delta E$, the most effective contributions are
those with the highest tunneling strength and lowest $\Delta E$.
This argument makes us only include the set of nearest ${\bf K}$ points.
Therefore, we diagonalize
\begin{equation}
 {\cal H}^{a} = \left[ \begin{array}{*2c} v({\bf q}-\alpha_a {\bf K}_{a0})_u + E_{RR}^{a} & E_{RL}^{a} \\ E_{LR}^{a} & -v({\bf q}-\alpha_a {\bf K}_{a0})_u+E_{LL}^{a} \\ \end{array} \right],
\end{equation}
where
\begin{equation}
E_{\lambda\lambda'}^{a} = \sum_{<\ell_a \ell_b>}\frac{ \Big[  T'(\alpha_a 
\lambda {\bf q}_a | \alpha_b L {\bf q}_b)  T'^*(\alpha_a \lambda' {\bf
q}_a | \alpha_b L {\bf q}_b) - T'(\alpha_a \lambda {\bf q}_a | \alpha_b
R {\bf q}_b)  T'^*(\alpha_a \lambda' {\bf q}_a | \alpha_b R {\bf q}_b)
\Big]}{v(\alpha_a {\bf K}_{a \ell_a} - \alpha_b  {\bf K}_{b \ell_b})_{u}}\Bigg|_{{\bf q}_a=0},
\end{equation}
and $\lambda,\lambda'=R,L$.

Diagonalizing, we get
\begin{equation}
E_{\pm}^{a}({\bf q}) =  \frac{E_{LL}^{a} + E_{RR}^{a}}{2}   \pm \sqrt{
\bigg(v({\bf q} - \alpha_a {\bf K}_{a0})_u -\frac{E_{LL}^{a} - E_{RR}^{a}}{2}  \bigg)
^{2} + |E_{RL}^{a}|^{2}}.
\end{equation}
The backscattering gap is $ E_{g}^{a} = 2|E_{RL}^{a}|$, and it
opens around an energy offset  $E_{o}^{a} = (E_{RR}^{a} +
E_{LL}^{a})/2$.  We thus find
\begin{equation}
E_{g}^{a} = {4 t_T^2 \over v K_0} \bigg| \sum_{<\ell_a \ell_b>} { e^{-{1\over 2} R a_0 K_0^2 (\alpha_a \sin\omega_{a \ell_a} + \alpha_b \sin\omega_{b \ell_b} )^2}  \over \alpha_a \cos\omega_{a \ell_a} - \alpha_b  \cos \omega_{b \ell_b}
}
 \sin\omega_{a \ell_a} \cos\omega_{b \ell_b}\bigg|,
\end{equation}
and
\begin{equation}
E_{o}^{a} = {2 t_T^2 \over v K_0} \sum_{<\ell_a \ell_b>} { e^{-{1\over 2} R a_0 K_0^2 (\alpha_a \sin\omega_{a \ell_a} + \alpha_b \sin\omega_{b \ell_b} )^2}  \over \alpha_a \cos\omega_{a \ell_a} - \alpha_b  \cos \omega_{b \ell_b}  }
\cos\omega_{b \ell_b}.
\end{equation}

Eqs(4.7) and (4.8) show that, in general, the gap offset $E_{o}$ is
greater than the gap $E_{g}$. In addition, one expects that the
offsets and gaps of both tubes will generally be different. This
means that the gaps wiggle around the Fermi energy as the chiral
angle is changed, leading to the conclusion that in a rope formed
of a random collection of chiralities, the effect on the density
of states around $E_{F}$ is very small.

Let us now have a closer look at the contributions of different tunneling points. In general, only one set of Fermi points will dominate, unless the tubes are mirror images of each other. We now argue that in a case when the tubes have different chiralities, it is a certain set of Fermi points that is actually important. 

We want to understand which tubes significantly couple to each other, and for those tubes, the Fermi points at which the coupling is most effective. To do this, we study the quantity
\begin{equation}
{t \over \Delta E} = 
{t_T \over v K_0} {e^{-{1 \over 4}R a_0 K_0^2 (\alpha_a \sin\omega_{a \ell_a} + \alpha_b \sin\omega_{b\ell_b} )^2} \over  \alpha_a \cos\omega_{a \ell_a} - \alpha_b \cos\omega_{b\ell_b} }.
\end{equation}
This quantity will be dominated by the exponential factor, and in cases where different sets of points have nearly equal exponential contribution, the denominator will dominate.

For nearly armchair tubes, the maximum is at $\omega_{a(b)\ell(\ell')} \sim 0$, i.e., around the ${\bf K}_0$ points(${\bf K}_0 {\mbox{-}} {\bf K}_0$ tunneling). In that case, the exponential factor is approximately $1$, and the denominator takes it minimum value($\sim 7$ meV) as the two $\omega$'s are closest to zero. As the tubes shift from being armchair, the denominator increases as the Fermi points rotate away from the $u$ axis, thereby making the tunneling less effective. 

For tubes which are nearly mirror images of each other, the dominant set is also the ${\bf K}_0 {\mbox{-}} {\bf K}_0$. While moving away from the armchair region does not  significantly change the exponential contribution, it makes the denominator bigger, hence making the coupling less important.

For tubes which are nearly zigzag(but are not mirror images of each other), tunneling is the least effective, as both the exponential argument and the denominator are big. In some of these cases, ${\bf K}_0 {\mbox{-}} {\bf K}_0$ tunneling  may not be the dominant one.

We thus conclude that tunneling is most effective between tubes which are nearly armchair, and that the ${\bf K}_0 {\mbox{-}} {\bf K}_0$ tunneling is the most important one, and hence $E_{RR}^{1}$(and similar sums) are dominated by one term.  In other words, the sum over reciprocal lattice vectors in the tunneling Hamiltonian is dominated by a single term.  If we ignore the other terms, then there is no
reciprocal lattice vector sum, and the system effectively has
translational invariance in the direction parallel to the tubes.
This ``dominant Fermi point approximation" simplifies our problem
considerably, since it allows us to assign a conserved momentum to
each state.  This will allow us to compute the band structure for
an entire rope in the following section.

\subsection{Compositionally disordered rope}

In this section we study the electronic structure of a nanotube rope composed of tubes with a random distribution of diameters and chiralities. We expect that the momentum mismatch between the Fermi points of neighboring tubes will suppress the tunneling and lead to localization. In real ropes, we expect $2/3$ of the tubes to be semiconducting. As indicated in Fig. 8, this will make the localization effects even stronger. To emphasize our point, we consider a compositionally disordered rope with only metallic tubes.

\begin{figure}
   \epsfxsize=3.0in
   \centerline{\epsffile{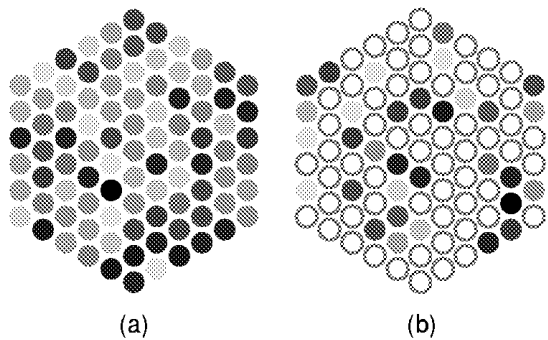}}
   \caption[FIG. 8.]{(a) A compositionally disordered metallic rope. Different gray scales indicate different tube chiralities. (b) A compositionally disordered rope with $1/3$ of its tubes metallic. The vacant circles denote semiconducting tubes. It is clear that the semiconducting tubes percolate along the rope.}
\end{figure}

To solve the problem, we employ the ``dominant Fermi point approximation'' introduced in the preceding section. In this approximation, the momentum $k = K_{0u} +  q$ is conserved by the tunneling Hamiltonian. We may thus write 
\begin{equation}
{\cal H} = \sum_{i} {\cal H}_{i} + \sum_{<ij>} {\cal H}_{ij},
\end{equation}
with
\begin{equation}
{\cal H}_i = \sum_{\alpha k} v( k - \alpha  K_{0u}^i)(\psi^{\dagger}_{i \alpha k R} \psi_{i \alpha k R} - \psi^{\dagger}_{i \alpha k L} \psi_{i \alpha k L}) ,
\end{equation}
and
\begin{equation}
{\cal H}_{ij} = \sum_{\alpha \eta'_i \eta'_j k} \widetilde{T'}(\alpha \eta'_i | \alpha \eta'_j )
\psi_{i\alpha k \eta'_i}^\dagger
\psi_{j\alpha k \eta'_j},
\end{equation}
where $\widetilde{T'}$is the tunneling matrix given by
\begin{equation}
\widetilde{T'} = 
 2 t_T e^{-{1\over 4} R a_0 K_0^2 ( \sin\omega_{i 0} + \sin\omega_{j 0} )^2}  \left[
        \begin{array}{*2c}
           \cos {\omega_{i 0} \over 2}  \cos {\omega_{j 0} \over 2}
      &  i \cos {\omega_{i 0} \over 2}  \sin {\omega_{j 0} \over 2}  \\
 	-i \sin {\omega_{i 0} \over 2}  \cos {\omega_{j 0} \over 2}
      &    \sin {\omega_{i 0} \over 2}  \sin {\omega_{j 0} \over 2}
        \end{array} \right].
\end{equation}

The Hamiltonian may now be diagonalized for each $k$ by diagonalizing a $4N \times 4N$ matrix, where $N$ is the number of tubes in the rope. For each $k$, the $m^{\mbox{th}}$ eigenstate  may be described by a ``wavefunction'' $\zeta^m_{\alpha \eta}(i)$, which is the amplitude for the particle to be in state $\alpha$, $\eta$ on tube $i$.

A portion of the band structure of the metallic rope is shown in Fig. 9a. It is clear that there is no significant change in the vicinity of  the Fermi energy. As we
have argued before the backscattering gaps wiggle around the Fermi
energy, thereby negligibly changing the density of states, which
is shown in Fig. 9b. Therefore, no pseudo gap develops.

\begin{figure}
   \epsfxsize=4.0in
   \centerline{\epsffile{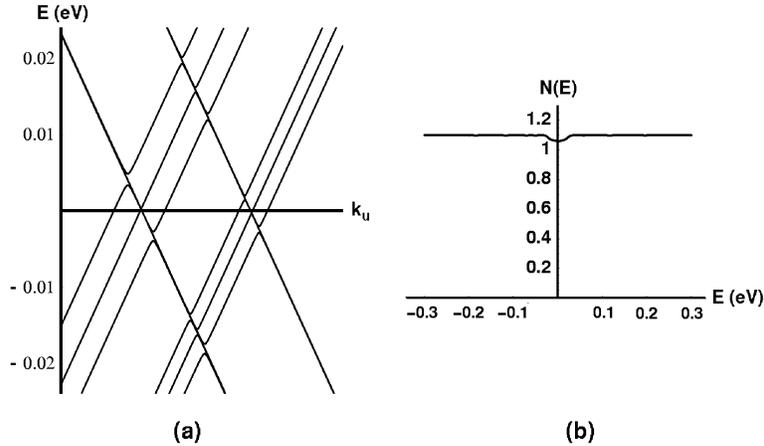}}
   \caption[FIG. 9.]{(a) Low energy band structure of a compositionally disordered metallic rope, showing tunneling and backscattering gaps. The latter wiggle around the Fermi energy, leading to a negligible effect on the density of states, which is shown in (b).}
\end{figure}

The extent of localization of the eigenstates may be quantitatively measured by computing the correlation function
\begin{equation}
C(r_\perp,E) = \sum_{i j m \alpha \eta \alpha' \eta'} | \zeta^m_{\alpha \eta}(i) |^2 | \zeta^m_{\alpha' \eta'}(j) |^2  \delta(|{\bf R}_i - {\bf R}_j | - r_\perp) \delta(E_m - E),
\end{equation}
at the Fermi energy, where ${\bf R}_i$ is the position of tube $i$ in the rope.
As shown in Fig. 10, the correlation function decays exponentially with distance, $C(r_\perp,E_F) \propto e^{-2r_\perp /\xi_\perp}$, indicating that the eigenstates are localized perpendicular to the tube axes with a localization length $\xi_\perp \sim 10{\mbox{\AA}}$. Thus the eigenstates are predominantly on a single tube.

\begin{figure}
   \epsfxsize=3.0in
   \centerline{\epsffile{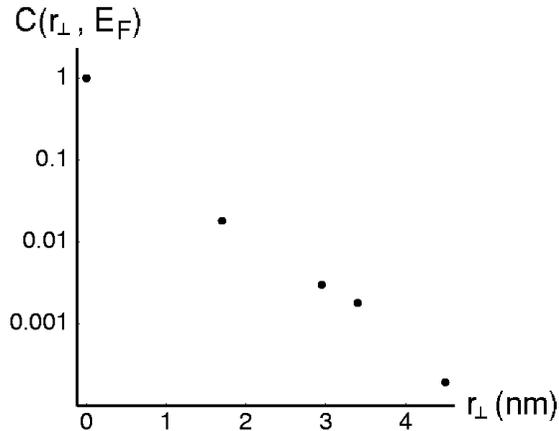}}
   \caption[FIG. 10.]{The correlation function $C(r_\perp,E_F)$ defined by eq(4.14), which gives a quantitative measure of the localization of the states on single tubes.}
\end{figure}

\section{Conclusion}

In this paper we have shown that the constraints of energy and
crystal momentum conservation severely restrict the electronic
coupling between carbon nanotubes.  The electronic coupling
between two nanotubes is only effective when the eigenstates near
the Fermi energy have the same momentum, which requires that the
graphene sheets of the two nanotubes are oriented parallel to one
another. This only occurs when the two tubes are mirror images of
one another.  Thus, in contrast to a crystalline rope of armchair
nanotubes, in which eigenstates are extended throughout the rope,
we find that the electronic eigenstates of a compositionally
disordered rope are strongly localized on individual nanotubes.

This conclusion has important consequences for the transport
properties of nanotube ropes.  In particular, it provides a
natural explanation of the nonlocal effects observed by Bockrath
et al. \cite{Bockrath} in their multi-terminal conductance measurements.  These
effects can arise when different electrical leads make contact to
different tubes within a rope, allowing the current in a tube to
``bypass" an electrical lead which it does not contact.

In the absence of impurities, the eigenstates will be localized on
a single tube, but extend across the entire length of a tube.
Scattering, either due to impurities or tube ends, will tend to
localize the states in the tube direction.  Paradoxically, by
relaxing the constraint of momentum conservation such scattering
will {\it increase} the coupling between tubes.  Nonetheless, we
are led to a picture of {\it highly anisotropic} localization.

This picture may help to explain some apparently paradoxical
transport data on nanotube mats.  At low temperatures, nanotube
mats are observed to obey the three dimensional Mott variable range
hopping law, $R = R_0 \exp (T_0/T)^{1/4}$, with $T_0$ of order
100 K \cite{Antonov}.  If one uses the standard formula for {\it
isotropic} variable range hopping and knowledge of the nanotube's
density of states, one extracts a localization length of order
$200${\AA}. By contrast, Fuhrer et al. \cite{Fuhrer} have analyzed the scaling of the
hopping conductivity with electric field and temperature $R(E,T) =
f( \xi E/T)$, and have argued that the localization length is much
longer, of order 6000{\AA}. Our picture of anisotropic localization
offers a possible resolution to this discrepancy.  In the simplest
model of anisotropic variable range hopping, $T_0$ depends on the
{\it geometric mean} of the localization lengths, $(\xi_\parallel
\xi_\perp^2)^{1/3}$, while the scaling with electric field depends
on the {\it longest} localization length, $\xi_\parallel$.

In this paper, we have developed a general framework for describing
the electronic coupling between graphene based structures.  This
approach should prove useful for other problems, including the
coupling between neighboring shells of multiwalled tubes as well
as the coupling between crossed single walled tubes.  Analysis of
these problems will be left for future work.

\appendix

\section{Lattice Fourier Transforms}

In this appendix, we work out explicitly the Fourier transform in
section IIA. As shown in Fig. 1 the positions of the lattice may be
written as
 ${\bf r}_i = {\bf R} + {\bf \rho} + \eta {\bf \tau}$, which may be
specified by a lattice vector ${\bf R}$, and a sublattice index
$\eta = \pm 1$.  The position of the center of the hexagon is
given by $\rho$.  Consider first a sum over lattice sites of the
form
\begin{equation}
{1\over \sqrt{N}}\sum_i f({\bf r}_i) e^{-i {\bf k}\cdot {\bf r}_i}
 ={1\over\sqrt{N}} \sum_{{\bf R}\eta} f({\bf r})e^{i {\bf k}\cdot {\bf
r}}
 |_{{\bf r} = {\bf R} + {\bf \rho} + \eta\tau}.
\end{equation}
The sum over ${\bf R}$ may be performed by introducing reciprocal
lattice vectors ${\bf G}$,
\begin{equation}
={1\over{A_{\rm cell}\sqrt{N}}} \sum_{{\bf G}\eta} \int d^2 r
e^{-i{\bf G}\cdot({\bf r} - \rho - \eta \tau)} f({\bf r}) e^{-i
{\bf k}\cdot {\bf r}}.
\end{equation}
Defining the Fourier transform,
\begin{equation}
f_{\bf k} = {1\over {A_{\rm cell}\sqrt{N}}} \int d^2r t({\bf r})
 e^{-i{\bf k}\cdot{\bf r}},
\end{equation}
we may then write
\begin{equation}
{1\over \sqrt{N}}\sum_i f({\bf r}_i) e^{-i {\bf k}\cdot {\bf r}_i}
= \sum_{{\bf G}\eta} e^{i{\bf G}\cdot {\bf d}_\eta} f_{{\bf
G}+{\bf k}}. \end{equation}

\end{document}